\documentclass[
10pt, 
letterpaper, 
oneside, 
headinclude,footinclude 
]{scrartcl}

\usepackage[nochapters]{classicthesis} 
\RequirePackage{xcolor} 
\definecolor{halfgray}{gray}{0.55} 
\definecolor{webgreen}{rgb}{0,.5,0}
\definecolor{webbrown}{rgb}{.6,0,0}

\usepackage{hyperref}
\usepackage[T1]{fontenc} 
\usepackage[utf8]{inputenc} 
\usepackage{graphicx} 
\usepackage{enumitem} 

\hypersetup{
colorlinks=true, breaklinks=true, bookmarks=true,bookmarksnumbered,
urlcolor=webbrown, linkcolor=webbrown, citecolor=webgreen, 
pdftitle={}, 
pdfauthor={\textcopyright}, 
pdfsubject={}, 
pdfkeywords={}, 
pdfcreator={pdfLaTeX}, 
pdfproducer={LaTeX with hyperref and ClassicThesis} 
}

\usepackage{setspace}
\usepackage{graphicx}
\usepackage{booktabs}
\usepackage{textcomp}
\usepackage{cite}
\usepackage{authblk}
\usepackage{array}
\usepackage{caption}

\title{\normalfont Under a cloud of uncertainty: Legal questions affecting Internet storage and transmission of copyright-protected video content}

\author{Fraida Fund}
\author{S. Amir Hosseini}
\author{Shivendra S. Panwar}
\affil{\small Department of Electrical and Computer Engineering \\ NYU Tandon School of Engineering}

\date{}

\begin{document}

\renewcommand{\sectionmark}[1]{\markright{\spacedlowsmallcaps{#1}}} 
\lehead{\mbox{\llap{\small\thepage\kern1em\color{halfgray} \vline}\color{halfgray}\hspace{0.5em}\rightmark\hfil}} 

\pagestyle{scrheadings} 

\maketitle 
{\let\thefootnote\relax\footnotetext{{
\noindent This paper is to appear in the IEEE Network 
Smart Data Pricing Special Issue, to be published in March 2016. \\
\copyright 2015 IEEE. Personal use of this material is permitted. Permission from IEEE must be
obtained for all other uses, in any current or future media, including
reprinting/republishing this material for advertising or promotional purposes, creating new
collective works, for resale or redistribution to servers or lists, or reuse of any copyrighted
component of this work in other works
}}}

\setcounter{tocdepth}{2} 




\begin{abstract}
The rapid growth of multimedia consumption has 
triggered technical, economic, and business 
innovations that improve the quality and accessibility of 
content. It has also opened new markets, promising large 
revenues for industry players. However, new technologies 
also pose new questions regarding the legal aspects of content delivery,
which are often resolved through litigation between copyright 
owners and content distributors.  
The precedents set by these cases will act as a game changer in the content 
delivery industry and will shape the existing offerings in the market 
in terms of how new technologies can be deployed and what kind of pricing strategies 
can be associated with them. 
In this paper, we offer a tutorial on key 
copyright and communications laws and decisions related to 
storage and transmission of video content over the Internet. 
We summarize legal limitations on the 
deployment of new technologies and pricing mechanisms, and explain 
the implications of recent lawsuits. 
Understanding these concerns is essential for engineers engaged in 
designing the technical and economic aspects of video delivery systems.
\end{abstract}

\section{Introduction}

In North America, real time entertainment constitutes almost 
69\% of peak period downstream traffic in fixed networks and 40\% in mobile networks.
Netflix alone accounts for more than 36\% of peak period downstream 
traffic in fixed networks, with YouTube, Amazon Video, and Hulu
also appearing among the top ten peak period applications~\cite{sandvine15}.
These services collectively are both the greatest stress on current networks
(and thus, the primary contributor to costs of data delivery), 
and the greatest driver of demand for network services (and thus, 
a key component of revenue strategy for network service providers.) 

Recent proposals to ease the impact 
of video traffic on networks include  
smart data pricing schemes~\cite{sdpsurvey} involving
shifting video traffic in time or space~\cite{time-shift,chiao2006video},
better content delivery networks~\cite{cdn}, 
proactive caching~\cite{chen2009caching}, and
peer to peer delivery~\cite{vod,vratonjic2007enabling}. 
Some regulatory aspects of network pricing are also well investigated
(such as network neutrality ~\cite{misra2015routing}.)
However, because most of the video 
traffic under consideration is copyright-protected, 
new techniques must overcome additional legal and regulatory hurdles
to be applied in practice.
These have been discussed
in law, economics, and policy 
forums,~\cite{sherman2014future,thierer2014video,saw2015whither} 
but largely neglected in the engineering literature.

\begin{figure}[t]
    \centering
        \includegraphics[width=\textwidth]{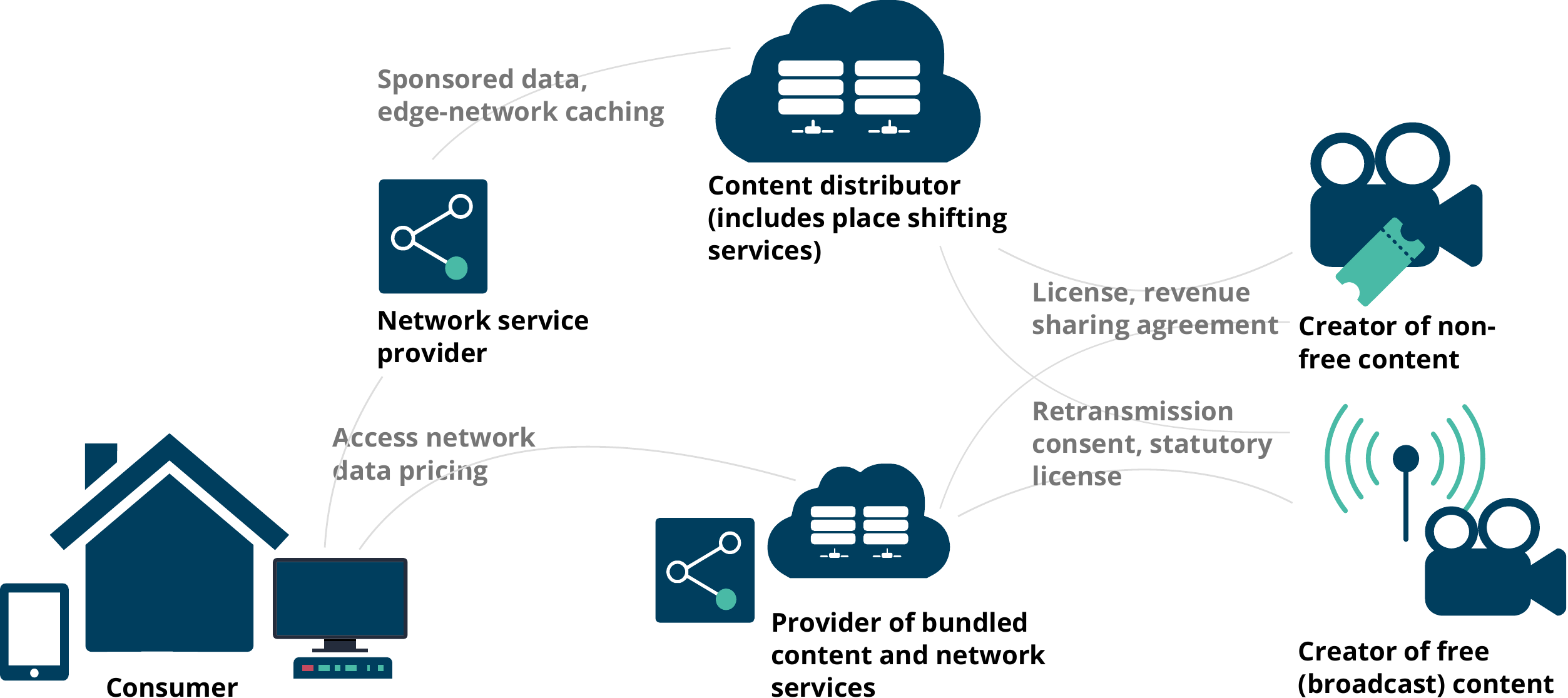}
    \caption{Delivery of multimedia content over the Internet involves financial
    and other relationships between several key players. The specific agreements
    between network service providers (Verizon, Comcast, AT\&T), content distributors (Netflix, Amazon), and content creators (Disney, Sony)
    affect both the availability and pricing of multimedia content and network services
    to consumers.}
    \label{fig:ecosystem}
\end{figure}

Key considerations for engineers designing technical protocols and pricing strategies
for Internet content delivery include: Where is content stored? Who initiates
the storage and/or transmission of content? What agreements exist
between the content creator, content distributor, and network service provider?
These questions are complicated by the diversity of the video content
delivered over the Internet, especially as the increasing population of 
``cord cutters'' opens a market for IPTV~\cite{iptv} or over-the-top 
delivery of traditional television content over the Internet.
The ecosystem of Internet video is going to include third-party licensed
video on demand, original programming included in bundles that include
broadband Internet and video service, broadcast TV content on demand, 
and live TV content. These, in turn, involve complex relationships
between content creators (Disney, Sony), content distributors (Netflix, Amazon), 
and network service providers (Verizon, Comcast, AT\&T)
as pictured in Figure~\ref{fig:ecosystem}, with associated legal challenges
that affect engineering and pricing strategies.

This tutorial considers issues related to the 
Internet distribution of copyrighted video content in the United States.
We begin with a brief overview of relevant copyright and communications 
legislation.
Then, we discuss selected legal challenges affecting technical and economic
solutions to the content delivery problem.
We conclude with a discussion of the implications for engineering
future multimedia content delivery networks. 

\section{Copyright and communications legislation in the United States}\label{sec:copyright}

In the United States, the delivery
of copyrighted video content over the Internet is subject to copyright and communications law.

With respect to copyright, the 1909 Copyright Act (Public Law 60-349, 35 Stat. 1075) 
confers six exclusive rights upon the owner 
of copyrighted material. These include the right to (1) reproduce the work, (2) prepare
derivative works, (3) distribute copies of the work,
(4) publicly perform the work, (5) display the work publicly, and (6)
perform a digital audio transmission publicly.
The legal questions surrounding Internet storage and transmission 
focus on the copyright holders' exclusive rights of reproduction and public performance.
The definitions of these terms have been revisited recently 
as new techniques are devised for sharing content.

Digital technology has complicated the concept of reproduction. 
A legal ``copy'' is one in which the work is ``fixed'' in a material object, 
by some method from which it can be reproduced or communicated. 
The advent of computer memory raises some questions regarding the 
definition of ``fixed''. Is a digital copy ``fixed''  if it is stored
in volatile memory (RAM)? What if it is stored
on a hard disk but the file descriptor is erased immediately?

Digital technology has also complicated the concept of public performance.
The original definition of ``public performance'' 
in the Copyright Act was very narrow.
In \emph{Teleprompter Corp. v. Columbia Broadcasting Systems, Inc.} (415  U.S.  394  (1974))
and \emph{Fortnightly Corp. v. United Artists Television, Inc.} (392  U.S.  390  (1968)),
the United States Supreme Court decided that transmission of broadcast 
television via cable did not constitute a ``public performance'' and thus
cable providers who retransmit broadcast television are  not infringing.
In response to these decisions, Congress amended the Copyright Act in 1976 (Public Law 94-553),
expanding the definition of a ``public performance'' to include the transmission of a work
to the public ``by means of any device or process, whether the members 
of the public capable of receiving the performance or display receive 
it in the same place or in separate places and at the same time or at different times.''
This prevented businesses (including cable providers) from selling access
to broadcasters' signals without compensation. 

At the same time, Congress implemented
a licensing arrangement to minimize the burden on 
cable providers, while still protecting broadcasters. By law, a 
service wishing to retransmit broadcast signals must negotiate a license from copyright providers
and also gain the consent of broadcasters. 
However, depending on its legal classification under communications law, 
a service may be eligible to participate in two regulated markets that ease the burden of negotiation.

First, Sections 111, 119, and 122 of the Copyright Act 
(U.S. Code Title 17, Chapter 1)
grant cable and satellite providers meeting certain requirements the right to retransmit broadcast 
television programming without negotiating with individual copyright holders.
With this ``compulsory license,'' the provider either pays 
set (regulated, below market rate) royalty fees which are collected by the United States Copyright Office and then 
distributed to the copyright holders, or is entitled to 
a royalty-free license.

Another set of laws apply to negotiations between the broadcaster
and the retransmitter.  
The Communications Act (CFR, Title 47, Chapter I, Subchapter C, Part 76, Subpart D, Section 76) 
requires businesses classified as multichannel video programming 
distributors (MVPDs) to get ``retransmission consent'' to retransmit broadcast
television signals. (This is distinct from the license agreement with the copyright holder.)
Retransmission consent may involve compensation from the retransmitter to the broadcaster. 
Alternatively, eligible television broadcast stations may elect not to require retransmission 
consent, instead participating in a ``must carry'' arrangement which under
some conditions may require a cable operator that 
serves the same market to carry its signal.

\begin{figure}[t]
    \centering
        \includegraphics[width=\textwidth]{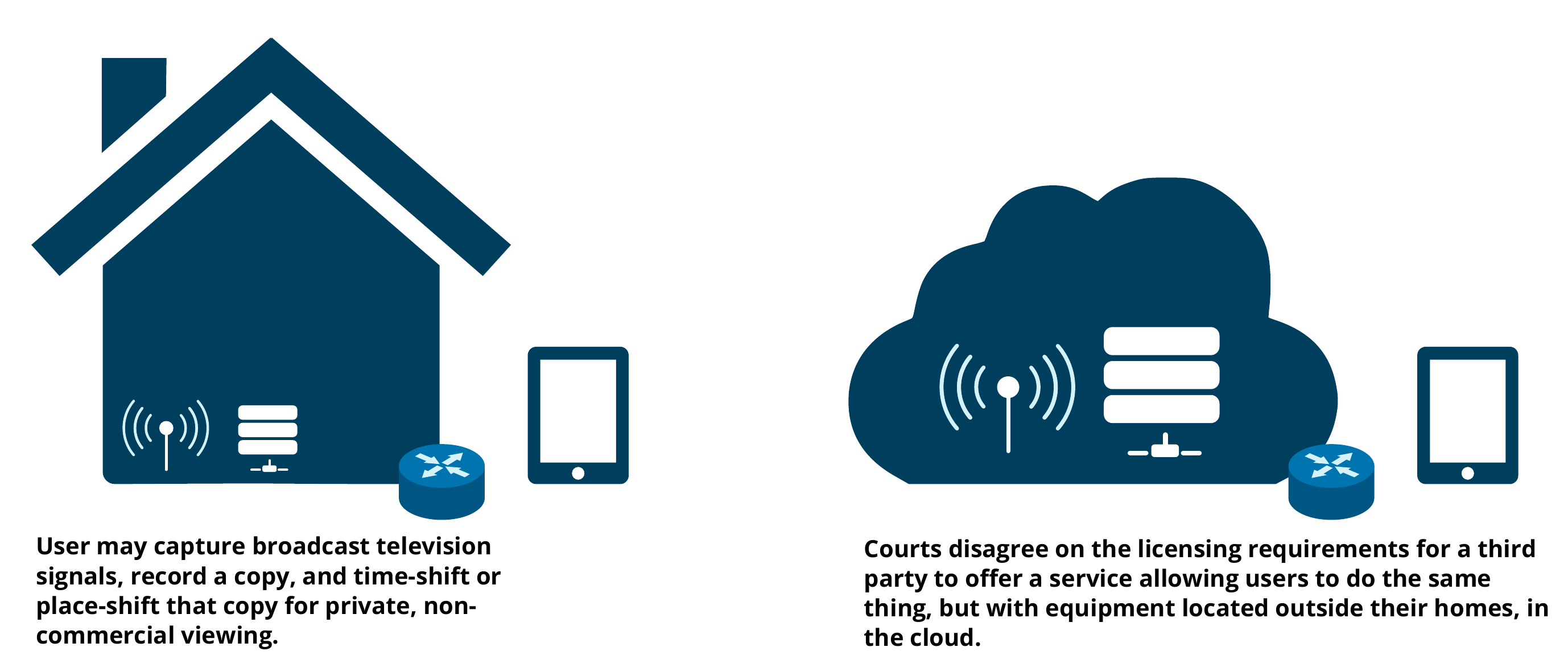}
    \caption{While the courts' interpretation of copyright and communications law
    allows consumers to record 
    video content for personal use in their own homes (\emph{Betamax})
    or stream video from their home set-top boxes to another device they own 
    (\emph{Fox Broadcasting Co. v. Dish Network, LLC}), the courts have been divided on
    the status of cloud services engaged in similar practices (\emph{Cablevision}, \emph{ivi}, \emph{Aereo}).}
    \label{fig:drawing}
\end{figure}

The implications of these laws for cloud services is not clear. While
case law is well established with respect to what end users or cable providers may do with video,
courts are divided on how to apply this to Internet-based services
(as in Figure~\ref{fig:drawing}).
Some approaches to smart data pricing
depend on cloud content providers' ability to freely shift data in 
time and/or space, and negotiate on fair terms with Internet service providers
that carry their content. The courts' interpretation of 
copyright and communications law for cloud providers
has a major impact on the licensing structures and costs
associated with each of these approaches.

\section{Time-shifting services, at home and in the cloud}

Many solutions to the content delivery problem propose time-shifting 
content delivery, in order to smooth traffic during 
peak periods~\cite{time-shift}. 
Smart pricing schemes may encourage users to modulate their viewing
habits and make delivery time more flexible for network service providers. 
However, it is unclear under what conditions a service engaging in 
this practice infringes the rights of the copyright holder (in which
case, it would require a potentially costly license 
that might negate the savings associated with time-shifting).

Given that the exclusive rights to reproduction and public performance are held by the 
copyright owner, is a broadband service provider allowed to store
copyrighted content for users in the ``cloud'' and deliver it to them later, on request?
The key precedents in the United States are the 1984 Supreme Court decision
in \emph{Sony Corp. of America v. Universal City Studios, Inc. (Betamax)} (464 U.S. 417 (1984))
and the 2008 Second Circuit decision in 
\emph{Cartoon Network LP v. CSC Holdings Inc. (Cablevision)} (536 F.3d 121 (2d Cir. 2008)).

\phantomsection
 \label{sec:betamax}
\subsubsection{Sony Corp. of America v. Universal City Studios, Inc. (Betamax)}

In 1979, Sony was sued by members of the film industry for 
its role in developing the Betamax VCR. The plaintiffs claimed 
that because Sony was manufacturing a device that could be used for copyright 
infringement, they were liable for infringement committed by its customers.
The Supreme Court decision found Sony not liable because the Betamax VCR
had non-infringing uses, and concluded that ``private, noncommercial time-shifting in the home''
is fair use and does not infringe on the reproduction right.

\emph{Betamax} addresses two key questions with implications for future cloud services:

\begin{itemize}
\item \emph{Can a company be held liable for infringement if the service 
it provides has both infringing and non-infringing uses?} 
The court decided that the Betamax VCR had significant non-infringing uses, 
for example making copies of televised content with permission of the copyright
holder, and that Sony was therefore not liable for potential infringement.
\item \emph{May a viewer time-shift video content without the authorization
of the copyright holder?} The court further decided that 
time-shifting television for private, non-commercial use in the home
is permitted even without the authorization of the copyright holder, 
as it ``merely enables a viewer to see such a work which he had been 
invited to witness in its entirety free of charge.''
\end{itemize}

Later legislation (notably the Digital Millenium Copyright Act) 
and case law modified the Sony decision in several ways.
In \emph{A\&M Records, Inc. v. Napster, Inc.} (239 F.3d 1004 (9th Cir. 2001)), an appeals court 
found that Napster \emph{could} be held liable for contributory infringement
because it was able to monitor and control users' activities. Similarly, 
in \emph{MGM Studios, Inc. v. Grokster, Ltd.} (545  U.S.  913  (2005)), the Supreme Court decided that 
Grokster could be liable for inducing copyright infringement 
(despite having non-infringing uses) because Grokster 
advertised and instructed users on engaging in infringement.

\phantomsection
 \label{sec:cablevision}
\subsubsection{Cartoon Network LP v. CSC Holdings Inc. (Cablevision)}

In 2008, a consortium of copyright holders sued
Cablevision for its Remote Storage DVR (RS-DVR) service. 
Cablevision routed the multimedia data stream going to subscribers through a 
Broadband Media Router (BMR), where it was buffered for at most 1.2 seconds while the system 
checked if any customers had requested a copy. 
If a subscriber had requested a particular program, it would be
stored on a portion of a hard disk allocated to that subscriber
in Cablevision's cloud data center, from which the subscriber 
could later view it.

The lawsuit alleged direct copyright infringement, 
excluding the topic of contributory copyright infringement. In turn, 
Cablevision waived any defense based on fair use.
Thus, the precedent set by \emph{Betamax} is 
largely orthogonal to \emph{Cablevision}.

The court sided with Cablevision, setting key precedents in three areas:

\begin{itemize}
\item \emph{A copy of ``transitory'' duration does not infringe on the 
reproduction right.} Because Cablevision held content in its BMR buffer
for no more than 1.2 seconds at any time, its buffer copy did not 
infringe. This is in contrast to the earlier \emph{MAI Systems Corp. v. Peak Computer, Inc.} (991  F.2d  511  (9th  Cir.  1993)),
where the Ninth Circuit found that a copy held in volatile memory \emph{did} infringe.
\item \emph{If a customer issues the command to copy directly to a copying service, 
then the customer (not the service) is liable for the copy.} 
Although the copy stored on the hard disk
\emph{was} a reproduction in the legal sense, the court 
agreed that the customer, not Cablevision, was responsible for making the copy.
This decision establishes \emph{volitional action} as a key element
shielding cloud providers from liability for their users' actions.
\item \emph{A system that transmits to a single subscriber using 
a single unique copy produced by that subscriber does not constitute 
a ``public'' performance.} 
The court made the determination of whether a performance
is ``public'' based on the audience of the \emph{particular copy} of the work.
\end{itemize}

This decision is important for establishing copyright liability
protection for cloud providers, but the precedent it sets is limited. 
The protection against liability established by \emph{Cablevision} is based
on the grounds that the user directly initiates the copy; 
this protection would not necessarily apply to 
systems that proactively fetch content without an 
explicit request from the user. 
Similarly, based on \emph{Cablevision}, a network service provider that caches popular 
content at the edge of a network might need to store one copy per user, or else negotiate
a license for public performance of the content.

\subsubsection{Impact of time-shifting on data pricing}

The case law established by \emph{Cablevision} and \emph{Betamax} 
mostly affects smart data pricing techniques that involve time-dependent pricing~\cite{time-shift}. These 
may be more or less problematic depending on their implementation:

\begin{itemize}
\item Pricing strategies that encourage users
to change their viewing habits by deferring \emph{consumption}
to off-peak times are not likely to be problematic. 
\item Pricing strategies that encourage users
to pre-fetch content in off-peak times may be problematic in some cases. 
\emph{Betamax} established that pre-fetching
is permitted for private, non-commercial use in the home, but not 
necessarily for other uses. Furthermore, if the service provider
creates a non-transitory copy that is not user-initiated in order 
to facilitate the technical process of pre-fetching,
the precedent set by \emph{Cablevision} protecting 
the service provider from liability may not apply.
\item Pricing strategies in which content 
is proactively pre-fetched (e.g. based on predictions of content 
that is likely to be of interest to the consumer) during off-peak times 
without the user initiating the download can be problematic, 
as the precedent set by \emph{Cablevision}
requires volitional action on the part of the user to protect the 
service provider from liability.
\end{itemize}

\section{``X with a long cord:'' Place-shifting services}\label{sec:placeshift}

While \emph{Betamax} and \emph{Cablevision} addressed the issue of time-shifting, 
\emph{Warner Bros. Entertainment Inc. v. WTV Systems, Inc.} (824 F.Supp.2d 1003 (2011))
and 
\emph{Fox Broadcasting Co. v. Dish Network, LLC} (723 F.3d (9th Cir. 2013))
concern the copyright implications of place-shifting.  
Place-shifting allows viewers to watch video content at a place of their choosing. 
Typically, these services present themselves as ``X with a long cord'': 
a DVD and DVD player rental attached to a long cord 
(Zediva in \emph{Warner Bros. Entertainment Inc. v. WTV Systems, Inc.}),
a television set-top box with a long cord 
(Slingbox in \emph{Fox Broadcasting Co. v. Dish Network, LLC}), 
or a television antenna with a long cord (ivi, Aereo), 
for example.

\phantomsection
 \label{sec:zediva}
\subsubsection{Warner Bros. Entertainment Inc. v. WTV Systems, Inc.} 

Zediva was a service offered in 2011
that allowed customers to watch movies online by streaming a signal over the Internet
from physical DVD players located in California. 
Customers who ``rented'' a DVD had exclusive access to that disk 
and a DVD player for up to four hours. Each 
disk could only be viewed by one customer at a time.

Because they did not negotiate streaming licenses, 
Zediva was able to offer new releases as soon as they were available on DVD, 
before they were licensed to streaming services such as Netflix.
Also, Zediva was able to undercut competitors; 
customers could rent a physical disk and DVD player for  \$1.99,
while licensed streaming video services at the time
charged between \$3.99 and \$5.99 for new releases. 

Zediva's defense argued that they were identical to a brick-and-mortar rental store, 
which is not required to negotiate licenses from copyright owners for 
post-purchase rentals (under the first sale doctrine). Because the first sale doctrine is a defense only 
for reproduction and distribution, not public performance, the case rested
on whether or not Zediva infringed on the public performance right.

The district court decided against Zediva, rejecting their first sale doctrine-based defense 
and noting that they were clearly operating 
a streaming service, not a DVD rental service. 
Noteworthy conclusions of the court were:

\begin{itemize}
\item Zediva's streaming signals were a ``public performance''
even though customers were using the DVDs at different times, implying that 
successive transmissions of a single copy to multiple viewers can be considered public 
performance. (In \emph{Cablevision}, each copy was only \emph{ever} viewed by one user.)
\item The ``length of the cable'' may be determinative in deciding whether 
copyright infringement occurred. Zediva considered itself analogous to ``playing back a
movie from a DVD with a very long cable attached,'' but the court held them 
liable for transmission because the videos were received ``beyond the place
from which they are sent.''
\end{itemize}

\phantomsection
\label{sec:dish}
\subsubsection{Fox Broadcasting Co. v. Dish Network, LLC}

In 2013, Dish implemented a service allowing subscribers to 
view content from their home set top boxes over the Internet, using a streaming server 
installed in the home (Slingbox).
This allows subscribers to view live, on-demand, or recorded content that they 
have access to at home, from any location.

Fox argued that Dish infringed on
the public performance right.  However, the district court found that 
because the service could only be used by subscribers to get access
to their own recordings (which were considered fair use, according to \emph{Betamax}), 
and because the reproduction and transmission actions were initiated 
by volitional action on the part of subscribers (as in \emph{Cablevision}), 
there was no direct infringement.
Furthermore, a user's transmission of programming from one place to another 
is not a public performance because the content is already in the subscriber's
possession, as is the equipment. Thus, Dish does not engage in contributory 
infringement by enabling this behavior.

This decision is significant, because while at face value, the Slingbox seems
like just another ``X with a long cord,'' here, the long cord carries content
between equipment already in the user's possession. Thus, a long cord connecting 
a user's device to the cloud is not equivalent to a long cord connecting two devices 
belonging to the same user across the Internet.

\subsubsection{Impact of place-shifting on data pricing}

Place-shifting affects smart data pricing strategies
by changing the dynamics of cost and demand in the ecosystem of Figure~\ref{fig:ecosystem}. 
The agreements between a place-shifting service and content creators 
(e.g. license agreements, revenue sharing contracts) affect the cost, 
value, and availability of the service. This in turn influences the 
prices of data services, as the network service provider acts as a platform 
for delivery of the content. 

 Place-shifting services in which content is stored in the cloud
can potentially reduce the acquisition and storage costs of the content distributor. 
These savings can trickle down and generate consumer surplus that may 
influence data purchasing decisions.  However, the legality of services
that do not specifically negotiate licenses for cloud streaming
(as in Zediva) is uncertain, especially when a single copy of the content 
is transmitted to multiple users (i.e., when the savings 
to the content distributor are greatest.)
Content distributors that negotiate a license in the face of this uncertainty
pass on higher licensing costs to consumers.

Place-shifting services that allow mobile users
to access content that they previously could only view at home (like Slingbox)
may shift demand from inexpensive home broadband networks
to relatively expensive cellular networks, partially negating  
smart data pricing strategies that rely on the 
ability to shift demand in the opposite 
direction. Again, the affordability of these services (and thus, their impact 
on data pricing) varies depending on whether or not 
place-shifting services must negotiate licenses, since those licensing costs
are typically passed on to consumers.

Finally, place-shifting also includes time-shifting, so  
considerations related to time-shifting and data pricing also apply.

\section{Schr\"{o}dinger's cable duck: Internet delivery of broadcast television}\label{sec:internet}

Even more than other forms of time-shifting and place-shifting,
the application of copyright law to cloud-based services that retransmit
broadcast television over the Internet has been confusing and contradictory. The key 
point of debate is the classification of these services under communications law.
This has been the subject of recent litigation involving two 
services, ivi and Aereo, 
which we briefly describe here. A related service called ``FilmOn'' 
is still enmeshed in active litigation.

\phantomsection
 \label{sec:ivi}
\subsubsection{WPIX, Inc. v. ivi, Inc.} 

ivi, Inc. was a cloud service that allowed subscribers to watch local broadcast TV from several U.S. cities
for a monthly fee of \$4.99 (with an option to also purchase
a recording service for additional \$0.99). It was sued by a group of 
copyright holders and broadcasters one week after beginning retransmissions.
A district court (765 F. Supp. 2d 594 (S.D.N.Y. 2011)) and an appeals court (No. 11-788 (2d Cir. 2012)) decided 
against ivi, forcing them to cease operations.

ivi argued that they should be classified as a cable system, 
making them eligible for the compulsory license 
under \S{}111 and freeing them from the requirement to negotiate 
with copyright holders. The court indicated that it is unclear based solely on the text 
of the Copyright Act whether ivi should be considered a cable system. 
Thus, in ruling against ivi, the court based its decision on the
following considerations:

\begin{itemize}
\item The intent of Congress in enacting \S{}111 was to 
improve access for communities that were underserved by broadcast signals.
Because Internet-based retransmission is not localized
and is not intended mainly to support remote areas, the court found
that Congress did not intend for Internet retransmission services
to be eligible for compulsory licenses under \S{}111.
\item The United States Copyright Office has not 
interpreted \S{}111's compulsory licenses to include
Internet retransmission, which they have said 
they consider to be ``vastly different'' from other retransmitters who are eligible.
\end{itemize}

\phantomsection
\label{sec:aereo}
\subsubsection{American Broadcasting Companies v. Aereo}

Aereo allowed subscribers to stream
broadcast television over the Internet for \$8/month. 
The creators of Aereo designed the service specifically
to avoid infringing on reproduction or public performance rights, 
using the precedent set by \emph{Cablevision}.
Aereo set up an ``antenna farm'' in a warehouse in New York.
Users of the service ``rented'' an individual antenna and were also offered a
VCR service, allowing them to store copies of television 
programs for later streaming.

As in \emph{Cablevision}, the users engaged in volitional conduct to create a copy, 
no non-transitory copies except for the users' were created, and an individual copy 
(an individual antenna) was dedicated to each user.
Aereo allowed users to access content they were already permitted to view for free over public airwaves.
Both the district court and appeals court sided with Aereo, citing
\emph{Cablevision} as precedent.

However, the Supreme Court decided in favor of the broadcasters (134 S. Ct. 2498, 2511 (2014)).
The court applied the ``duck test'' (if it looks like a duck, swims like a duck, 
and quacks like a duck, then it probably is a duck), arguing that 
Aereo had an ``overwhelming likeness to cable companies'' and therefore, 
required the consent of broadcasters to retransmit 
their signals.

Following this decision, the company argued that since 
they are a cable system, they are eligible for the compulsory copyright license. 
This argument was rejected by the district court (Civil Action No. 12-CV-1540 (AJN) (HBP) (S.D.N.Y Oct. 29, 2014)), 
which called it a ``fallacy'' that ``because an entity performs copyrighted works 
in a way similar to cable systems it must then be deemed a cable system for all 
other purposes of the Copyright Act.'' Thus Aereo became Schr\"{o}dinger's cable duck: 
simultaneously a cable system and not a cable system. Although it was required, like a cable
system, to obtain retransmission consent from broadcasters, it was not 
eligible for compulsory licenses from copyright holders.

A similar service, FilmOn, has been the subject of more recent, but equally contradictory, 
litigation. In July 2015, a Los Angeles federal district court decided that FilmOn was entitled 
to the same compulsory licenses as cable companies. However, in November 2015, 
this judgment was rejected on appeal.
The ongoing FilmOn litigation highlights how vague the Aereo ruling was, 
and the level of uncertainty that remains regarding retransmission of
broadcast television.

\subsubsection{Impact of Internet retransmission of television on data pricing}

\begin{figure}[t]
    \centering
        \includegraphics[width=0.8\textwidth]{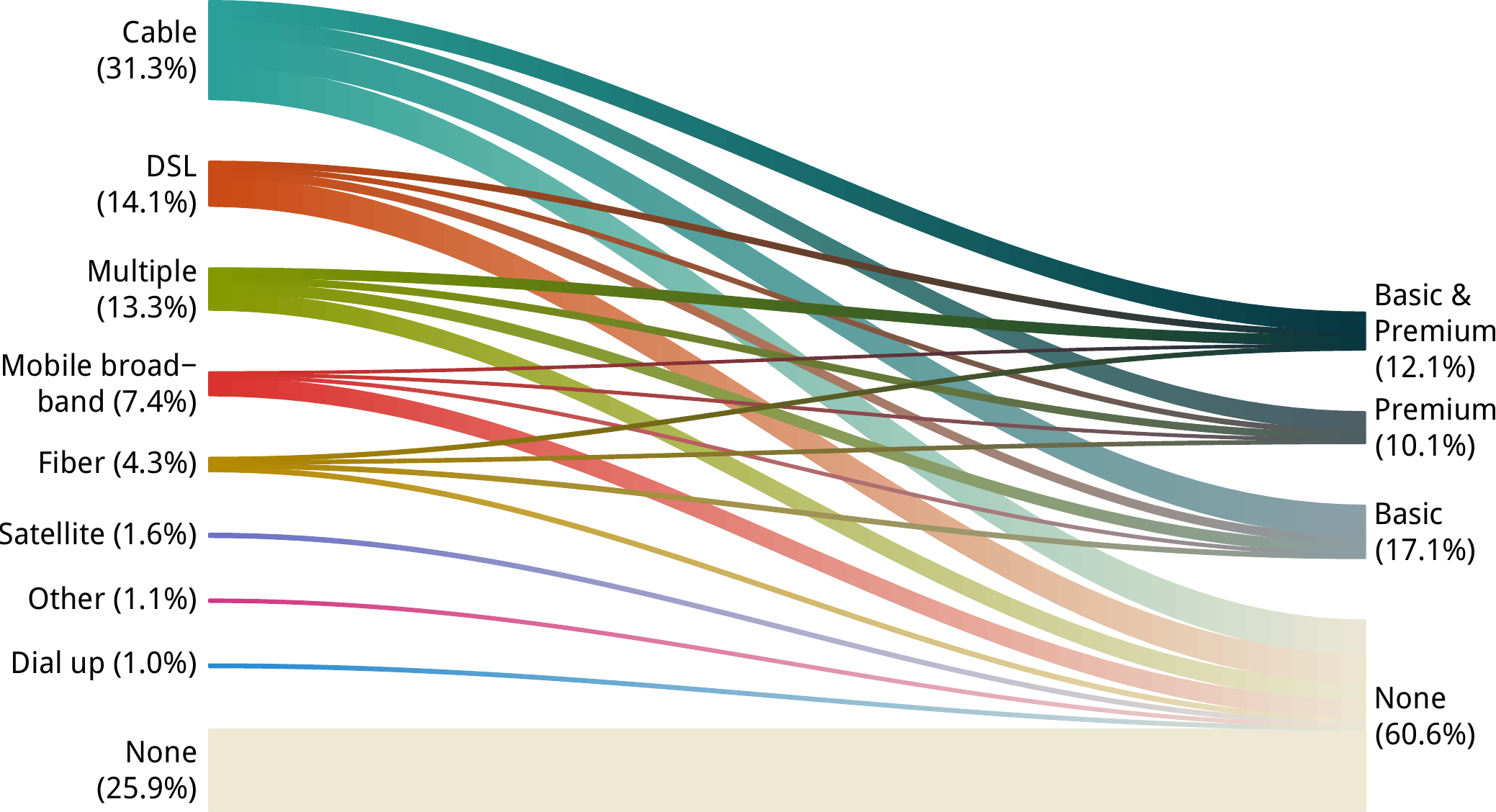}
    \caption{Close to 40\% of U.S. households purchase Internet service as part of a bundle 
    that also includes television services. On the left, we see the percentage of U.S. households
    who access the Internet using only cable, DSL, mobile broadband, fiber, satellite, multiple technologies, or
    who do not access the Internet from home. Moving towards the right, we see which of those purchase 
    Internet service as part of a bundle including basic and/or premium television services. 
    (Data source: July 2013 United States Census Computer and Internet Use Supplement. Percentages may not sum to 100 due to rounding.)}
    \label{fig:census_river}
\end{figure}

Smart data pricing often involves side payments and agreements between 
Internet-based content providers and network service providers (for example, 
as in sponsored data~\cite{zhang2014sponsoring}) under the assumption that these 
are separate parties with distinct interests. 
However, network service providers may have a competing interest 
in the content delivery market; many also sell video entertainment services. 
Figure~\ref{fig:census_river} 
shows that according to a July 2013
census survey,  39\% of U.S. households buy Internet service as part of a ``bundle'' 
including television. 

\begin{figure}[t]
    \centering
        \includegraphics[width=0.7\textwidth]{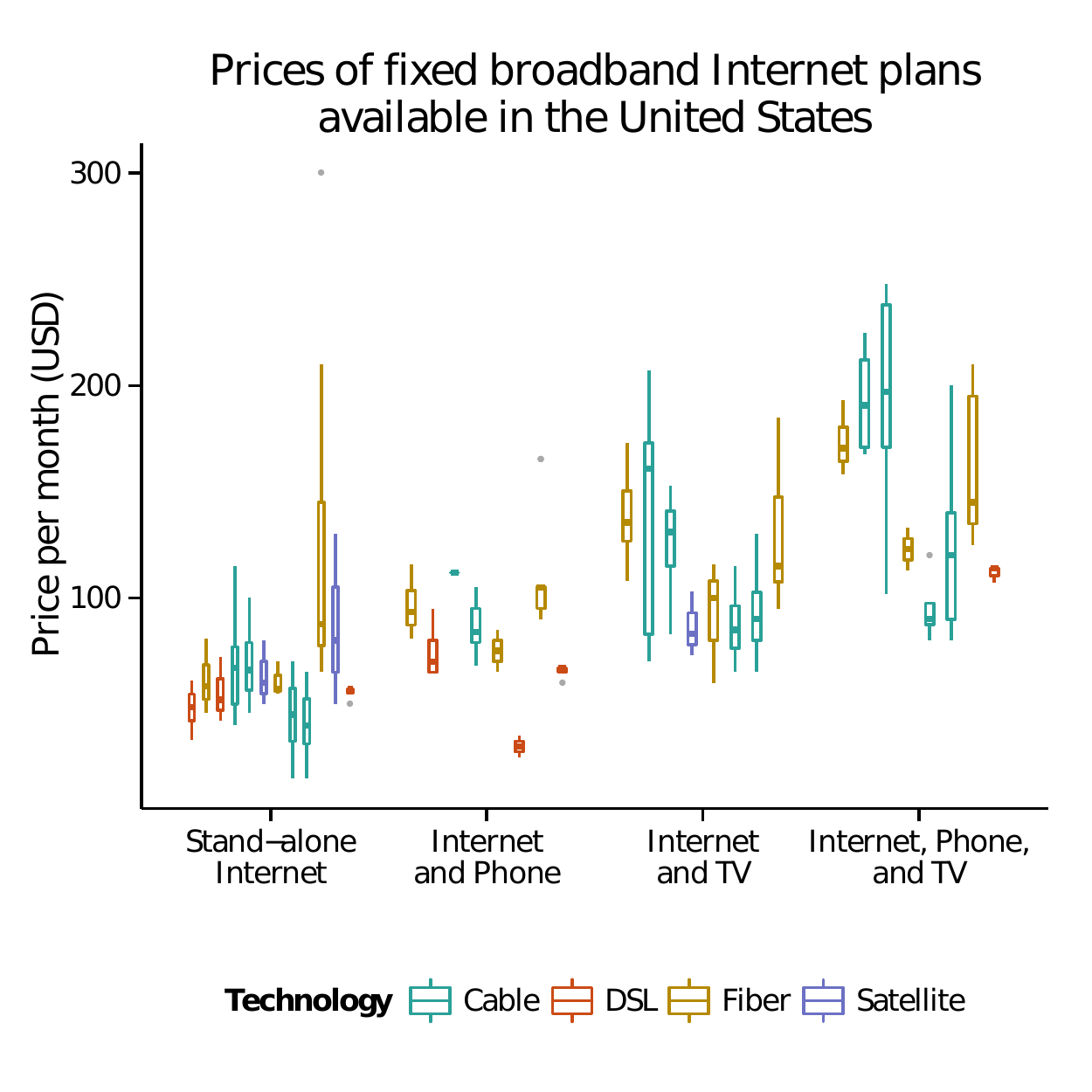}
    \caption{Prices for Internet service in the United States, grouped by technology, 
    bundle type, and provider. The lower and upper ``hinges'' of the boxplot correspond 
    to the first and third quartiles of available plans, the ``whiskers'' extend to 
    $1.5\times$ the interquartile range, 
    and outliers beyond the whiskers are plotted as points.
    A given provider may offer a range of plans of the same 
    bundle type using the same technology, differentiated by data rate, cap, or number of 
    television channels included. Not all plans are available in all U.S. markets. 
    (Data Source: International Bureau, Fourth International Broadband Data Report, 2015.)}
    \label{fig:prices}
\end{figure}

Bundles are important to cable operators, who have been losing pay-TV subscribers 
while the high-speed Internet user base continues to grow.
The cost of an Internet service bundle including other services is
higher than an equivalent stand-alone Internet plan (Figure~\ref{fig:prices}).
However, consumers have an increasing 
preference for \`a la carte television options or ``cord cutting.'' 
According to recent Nielsen reports, the number of ``zero TV'' households is on the 
rise, and almost half of ``zero TV'' households are composed of young people 
under the age of 35. Meanwhile, broadcast television networks remain important to 
viewers, with the ``Big Four'' still retaining a 40\% average weekly reach. 
The coincidence of these trends makes alternative video platforms, 
and Internet delivery of television programming in particular, 
the next big front in the battle for consumer dollars.

The ecosystem shown in Figure~\ref{fig:ecosystem} currently favors 
cable and satellite providers that also sell bundles including broadcast television, since 
these are entitled to compulsory licenses below market rates.
These services may even have a competitive edge when selling IPTV service 
(i.e., not traditional cable television). By delivering this service over managed IP, 
not the public Internet, they can offer a better quality of service than competitors and also 
exempt their own service from data caps, without running
into network neutrality issues that apply on public Internet.
(See for example Comcast's ``Stream'' product, announced July 2015.)

However, if services such as ivi, Aereo, and FilmOn were reclassified 
to also entitle them to compulsory licenses, the 
balance of this ecosystem would shift dramatically. This would affect 
aspects of data pricing related to transactions between 
content distributors and network service providers, including network neutrality~\cite{misra2015routing}, 
sponsored data~\cite{zhang2014sponsoring}, and app-based pricing~\cite{sdpsurvey}.

The Federal Communications Commission (FCC) and/or Congress are likely to 
eventually regulate Internet streaming providers, and describe what rights 
and requirements apply to them. It is not clear what this regulation might look like, 
and whether Internet streaming providers will at that stage be on equal footing 
with cable providers, who are subject to regulations imposed at a time when 
cable services operated mainly to improve access for underserved communities.

\section{Summary and global outlook}\label{sec:implications}

Table~\ref{tab:cases} summarizes key precedents 
set by cases discussed in this tutorial.
Given these decisions, the only certainty is that cloud multimedia services 
remain under a cloud of uncertainty.
This is a barrier to technical innovation, 
as without legal certainty regarding licensing requirements, 
companies offering new services are not able to predict costs, and investment in 
them is risky.

\noindent \begin{table}
\linespread{1}\selectfont\centering
\begin{tabular}{p{0.5cm}>{\raggedright}p{3.7cm}p{6.4cm}} \toprule
     & \multicolumn{2}{l} {\textbf{Time-shifting}} \\ \midrule
    1984  & \hyperref[sec:betamax]{Sony Corp. of America v. Universal City Studios, Inc. (Betamax)} 
 & ``Private, noncommercial time-shifting in the home'' 
    does not infringe on the reproduction right. \\  
    2008  & \hyperref[sec:cablevision]{Cartoon Network LP v. CSC Holdings Inc. (Cablevision)} & Establishes protection from liability if the service provider's copy is transitory and the 
user's copy is created by volitional action on the user's part. \\ \midrule
     & \multicolumn{2}{l} {\textbf{Place-shifting}} \\ \midrule
    2011  & \hyperref[sec:zediva]{Warner Bros. Entertainment Inc. v. WTV Systems, Inc.} & The ``length of the cable'' between the consumer and the content 
may be determinative in deciding whether copyright infringement occurs. \\
    2013 & \hyperref[sec:dish]{Fox Broadcasting Co. v. Dish Network, LLC} & A user's transmission of programming from one place to another
does not infringe on the public performance right when all content and equipment are in the subscriber's
posession.  \\  \midrule
     & \multicolumn{2}{l} {\textbf{Internet delivery of broadcast television}} \\ \midrule
    2012 & \hyperref[sec:ivi]{WPIX, Inc. v. ivi, Inc.} &  Internet-based broadcast television service ivi did not qualify for compulsory licenses to retransmit broadcast television. \\
    2014 & \hyperref[sec:aereo]{American Broadcast- ing Companies v. Aereo} & Aereo's place-shifting service required retransmission consent because they appeared too much like a cable provider, but another court judged that they did not resemble a cable provider enough to qualify for compulsory licenses.  \\  \bottomrule
\end{tabular}
\caption{Highlights of significant judicial decisions in the United States
related to time-shifting, place-shifting, and Internet delivery of broadcast television.}
    \label{tab:cases}
\end{table}

The global outlook for cloud multimedia services is similarly uncertain. 
Table~\ref{tab:global} shows cloud-based multimedia services that have been the subject 
of litigation (often dragging on for years, at considerable expense) in Europe and Asia.
These, too, have often yielded contradictory and vague decisions.

\noindent \begin{table}
\linespread{1}\selectfont\centering
\begin{tabular}{p{1.8cm}p{2.4cm}p{6.3cm}} \toprule
   \textbf{Location} & \textbf{Service} & \textbf{Status} \\ \midrule
    France & Wizzgo & Decided in favor of content owner \\  
    United Kingdom & TV Catchup & Mixed decisions, with litigation ongoing \\ 
    Finland & TVkaista & Decided in favor of content owner \\ 
    Germany & Shift.TV, Save.TV & Mixed decisions, with litigation ongoing \\  
    Singapore & Record TV & Decided in favor of cloud service \\ 
    Japan & Rokuraku II & Decided in favor of content owner \\ 
    Australia & Optus TV Now & Decided in favor of content owner \\  \bottomrule 
\end{tabular}
\caption{The global outlook for cloud multimedia services is equally uncertain. These services have been 
involved in extensive litigation, often with multiple appeals dragging on for years.}
    \label{tab:global}
\end{table}

The decisions made by legislative bodies, judicial bodies, and regulatory agencies
in the next few years as they apply copyright and communications law to the Internet will 
shape market offerings in this area, in terms of how new technologies can be deployed and what 
kind of pricing strategies can be associated with them.
Until then, however, key questions - 
when and where will network users consume multimedia content? what kinds
of relationships will exist between network service providers, 
content distributors, and content owners? - remain unanswered.

\section*{Acknowledgments}

This work was supported in part by the U.S. National Science
Foundation through the Graduate Research Fellowship Program Award
1104522, and by the New York State Center for Advanced
Technology in Telecommunications (CATT).

\end{document}